# Mining Characteristics of Vulnerable Smart Contracts Across Lifecycle Stages

**Hongli Peng[1], Xiaoqi Li[1*], and Wenkai Li[1]**
[1] School of Cyberspace Security, Hainan University
Haikou, Hainan 570228 China
[e-mail: csxqli@ieee.org]

## Abstract

Smart contracts are the cornerstone of decentralized applications and financial protocols, which extend the application of digital currency transactions. The applications and financial protocols introduce significant security challenges, resulting in substantial economic losses. Existing solutions predominantly focus on code vulnerabilities within smart contracts, accounting for only 50% of security incidents. Therefore, a more comprehensive study of security issues related to smart contracts is imperative. The existing empirical research realizes the static analysis of smart contracts from the perspective of the lifecycle and gives the corresponding measures for each stage. However, they lack the characteristic analysis of vulnerabilities in each stage and the distinction between the vulnerabilities. In this paper, we present the first empirical study on the security of smart contracts throughout their lifecycle, including deployment and execution, upgrade, and destruction stages. It delves into the security issues at each stage and provides at least seven feature descriptions. Finally, utilizing these seven features, five machine-learning classification models are used to identify vulnerabilities at different stages. The classification results reveal that vulnerable contracts exhibit distinct transaction features and ego network properties at various stages.

## 1. Introduction

Since the proposal of Bitcoin in 2008, blockchain technology has garnered significant attention for its diverse applications [1]. Blockchain operates as a decentralized and public ledger, eliminating the necessity for trusted third parties in financial systems by ensuring transparent and tamper-resistant data records. The advent of smart contracts ushered in the Blockchain 2.0 era, extending blockchain's applications beyond digital currency transactions to complex decentralized applications (DApps) and decentralized finance (DeFi) protocols. Today, smart contract-based blockchain technology is widely adopted across various industries, including healthcare [2–4], the Internet of Things (IoT) [5–8], and finance [9, 10]. This study specifically focuses on the Ethereum platform.

Despite their significant value and widespread adoption, smart contracts continue to face numerous security threats [11-16]. Research indicates that while some attacks directly target the code and can be mitigated through rigorous security checks before deployment and execution [17-25], many occur during other lifecycle stages [11, 26-31], such as upgrade and destruction. These stages pose unique challenges, including state transitions and governance oversight, often overlooked by traditional security methods. This lack of research highlights the insufficiency of ensuring comprehensive security across the smart contract lifecycle. Addressing these dynamic changes, particularly in the complex upgrade and destruction stages, is crucial for enhancing security.

To address these challenges, this paper presents an empirical study identifying 840 smart contracts with vulnerabilities across different lifecycle stages, using the SmartBugsV2 detection tool [32]. A comprehensive dataset of all related transactions is compiled via the Ethereum browser Etherscan [33]. Based on this dataset, we exhaustively examine security issues across the entire lifecycle, analyzing dynamic transactions and ego network characteristics during deployment and execution, upgrade, and destruction stages. Our analysis reveals distinct features of these vulnerable contracts, expanding the scope of smart contract security. Moreover, features associated with lifecycle stages are effectively distinguished from normal contracts using various machine learning models.

The main contributions of this paper are as follows:

(1) To the best of our knowledge, we present the first empirical study on the security of smart contracts throughout their lifecycle, including deployment and execution, upgrade, and destruction stages. By introducing feature descriptions of each stage and analyzing them from dynamic and ego network perspectives, we enhance our understanding of security challenges and facilitate the development of targeted mitigation strategies.

(2) The lifecycle of smart contracts is divided into three stages: deployment and execution, upgrade, and destruction. A systematic analysis examines security issues at each stage, providing a comprehensive perspective to explore vulnerabilities unique to each phase.

(3) We create a dataset comprising over *1.54 million* transactions involving more than 700 vulnerable smart contracts, categorized based on their lifecycle stages. Related codes and experimental data have been open-sourced at http://doi.org/10.6084/m9.figshare.24792717.

## 2. Background

### 2.1 Smart Contract Transactions

Smart contract transactions are the core operations that execute predefined code on the blockchain, enabling DApps and DeFi protocols. Each transaction is recorded on the blockchain, ensuring transparency and immutability. These transactions can include deploying contracts, interacting with existing contracts, or transferring assets. While enabling trustless automation, they are also vulnerable to risks if the contract code contains bugs or vulnerabilities.

### 2.2 Critical Opcodes in Smart Contract

DELEGATECALL and SELFDESTRUCT are critical opcodes in smart contracts. DELEGATECALL, a specialized message invocation in Solidity, allows code at a target address to execute within the context of the calling contract, retaining the original contract's storage, 'msg.sender' and 'msg.value' [34]. This opcode is pivotal during upgrades, enabling execution delegation to an updated contract without altering the original contract's state. It is also versatile, supporting modular architectures with dynamic interactions. In contrast, SELFDESTRUCT removes a contract from the blockchain and transfers its remaining balance to a designated address, crucial for controlled termination when contracts are obsolete or pose security risks. However, improper use of

SELFDESTRUCT can result in premature termination, leading to potential unintended consequences.

## 2.3 The Ego Network

The ego network refers to a subgraph within a larger network that includes a specific node, the Ego, and all nodes directly connected to it, known as Alters. The edges represent interactions between the ego and its first-order neighbors, as shown in **Fig. 1**. Ego network analysis is applied across various domains [35, 36], enabling researchers to explore individual interactions, information flow, and sub-group structures. This analysis provides insights into the roles and relationships of individuals, enhancing the understanding of characteristics inherent to each individual.

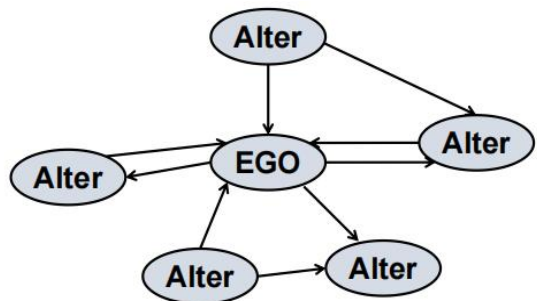

**Fig. 1.** An Example of Ego Network

## 3. Common Vulnerabilities in Various Lifecycle Stages

The entire lifecycle of smart contracts consists of three stages: deployment and execution, upgrade, and destruction. Initially, the contract is deployed to the blockchain, becoming operational and executing its encoded rules. Over time, upgrades may be needed to enhance functionality or address issues, often involving proxy contracts or opcodes like DELEGATECALL. The lifecycle concludes with the contract's destruction using the SELFDESTRUCT opcode, removing it from the blockchain once its purpose is fulfilled or termination is required.

Identifying the stages of a smart contract's lifecycle is challenging. DELEGATECALL indicates the upgrade stage, signaling vulnerabilities in the old contract and the need for a new version with similar functionality. SELFDESTRUCT marks the destruction stage, highlighting either vulnerabilities or task completion, prompting contract termination and asset transfer. Other vulnerabilities typically emerge during the deployment and operational stages.

Therefore, attention should be focused on DELEGATECALL-related vulnerabilities during the upgrade stage and SELFDESTRUCT-related vulnerabilities during the destruction stage. The remaining vulnerabilities primarily arise during the deployment and execution stages.

## 3.1 Common Vulnerabilities in Deployment and Execution Stage

Based on existing research [37], the common security vulnerabilities with high impact in smart contracts, excluding those related to DELEGATECALL and SELFDESTRUCT.

### 3.1.1 Integer Overflow and Underflow

The integer overflow or underflow vulnerability occurs when operations exceed the finite numerical range of integer types, causing the computer to truncate the result to fit the data type's limits. Attackers exploit this vulnerability for unauthorized actions, such as bypassing authentication or tampering with data [38]. Using validated secure mathematical libraries for arithmetic operations ensures proper handling of boundary cases, mitigating the risk of overflow and underflow.

### 3.1.2 Reentrancy

The reentrancy vulnerability, highlighted during the DAO attack [39], occurs when one smart contract invokes another without adequate controls, potentially leading to a reentrant loop. An attacker can repeatedly call functions during execution until a specified condition or computational resources are exhausted. To address this vulnerability, internal state changes should be completed before call execution, following the Checks-Effects-Interactions pattern. Reentrancy locks, such as ReentrancyGuard in the OpenZeppelin, can prevent recursive calls during external contract invocation and enhance protection.

### 3.1.3 Authorization through Tx.origin

The authorization through tx.origin vulnerability in Solidity arises from using the global variable tx.origin for smart contract authorization, exposing the transaction initiator's address and posing a security threat. To mitigate this risk, developers should avoid using tx.origin for authorization and instead use 'msg.sender', which accurately represents the immediate caller and is unaffected by external calls.

### 3.1.4 Timestamp Dependence

This vulnerability arises from smart contracts relying on timestamps, where attackers can manipulate transaction timestamps to extend or shorten time windows, bypassing restrictions and executing illicit operations [40]. To mitigate this, developers should minimize reliance on timestamps and use more reliable methods for contracts with time-related functionalities and constraints.

### 3.1.5 Unchecked Return Value

The unchecked return value vulnerability in smart contracts arises from the lack of checks on message call return values. If an exception is thrown by the called contract, execution continues without interruption, leading to potential unexpected behavior in subsequent operations [41]. Therefore, it is essential to carefully examine return values to address potential call failures.

### 3.1.6 Unprotected Ether Withdrawal

The unchecked return value vulnerability arises from a lack of checks on the return value of message calls in a smart contract, which may also cause unexpected behaviors in subsequent program execution [42]. Therefore, it is imperative to handle the potential for call failures by diligently examining the return values.

### 3.1.7 Assert Violation

The assert violation vulnerability arises from misusing the assert() function, which is intended for asserting invariants. Assertions should not fail under normal circumstances, and their failure may indicate a contract error or improper use, such as for input validation. To mitigate this, require() should be used for conditions essential to contract execution, while assert() should be reserved for checking conditions that should never be false in a properly functioning contract.

### 3.1.8 DoS with Failed Call

The DoS with failed call vulnerability occurs when an external call fails, potentially causing a DoS state. To mitigate this vulnerability, combining multiple calls in a single transaction should be avoided, particularly within loops.

## 3.2 Common Vulnerabilities in Upgrade Stage

### 3.2.1 DELEGATECALL to Untrusted Callee

The DELEGATECALL to untrusted callee vulnerability arises when a smart contract invokes another potentially untrusted contract using DELEGATECALL. This allows code execution at the target address within the current contract's context, preserving 'msg.sender' and 'msg.value'. Consequently, the caller is at risk of the invoked contract altering storage values and gaining control over the caller's balance. The use of DELEGATECALL requires careful consideration of associated risks, as untrusted contracts can exploit this functionality, jeopardizing the integrity of the state and financial holdings of the calling contract.

The code sample in **Fig. 2** demonstrates a vulnerability where the forward function can invoke any contract, including untrusted ones, without validating the address. This exposes the caller to potential manipulation of storage values and balance control by a malicious contract through DELEGATECALL.

To mitigate this risk, caution is advised when using DELEGATECALL, ensuring that untrusted contracts are not invoked. When the target address comes from user input, it is crucial to compare it against a trusted contract whitelist for enhanced security.

```solidity
1 contract Proxy {
2   address owner;
3   // Vulnerability 1: Missing contract
4   mapping(address => bool) trustedContracts;   whitelist
5   constructor() public {
6     owner = msg.sender;
7   }
8   function addTrustedContract(address _contract) public {
9     require(msg.sender == owner);
10    trustedContracts[_contract] = true;
11  }
12  function forward(address callee, bytes _data) public {
13    // Vulnerability 2: Lack of validation for the contract address
14    require(trustedContracts[callee]);
15    require(callee.delegatecall(_data));
16  }
17 }
```

**Fig. 2.** Example Code of The DELEGATECALL to Untrusted Callee Vulnerability

### 3.2.2 Payable Functions Using DELEGATECALL Inside A Loop

The payable functions using DELEGATECALL inside a loop vulnerability represents a security flaw in smart contracts, involving the utilization of DELEGATECALL within a loop structure in payable functions. This susceptibility may result in the repetitive accumulation of the same 'msg.value' amount within the loop, leading to inaccurate balance calculations.

The example code in Fig. 3 demonstrates a vulnerability where the bad function uses looped DELEGATECALL to pass addresses from the receiver array to the addBalance function. As addBalance is payable and utilizes 'msg.value' to increase address balances, a problem arises. The repeated passing of the same 'msg.value' within the loop leads to multiple increments of the address's balance, resulting in inaccurate calculations.

To address this, ensure that the called function avoids using 'msg.value' to prevent incorrect balance calculations caused by multiple invocations of the same 'msg.value' during delegatecall loops.

```solidity
1 contract DelegatecallInLoop {
2   mapping (address => uint256) balances;
3   function bad(address[] memory receivers) public payable {
4     for (uint256 i = 0; i < receivers.length; i++) {
5       // Vulnerability 1: Using DELEGATECALL in a loop with msg.value
6       address(this).delegatecall(abi.encodeWithSignature("addBalance(
          address)", receivers[i]));
7     }
8   }
9   function addBalance(address a) public payable {
10    // Vulnerability 2: Accumulating msg.value without proper checks
11    balances[a] += msg.value;
12  }
13 }
```

**Fig. 3.** Example Code of The Payable Functions Using DELEGATECALL Inside A Loop Vulnerability

### 3.3 Common Vulnerability in Destruction Stage

The vulnerability known as unprotected SELFDESTRUCT refers to inadequately secured invocations within a smart contract that can trigger the contract's self-destruct functionality. The SELFDESTRUCT operation could be exploited maliciously, leading to the unauthorized destruction of the contract and, consequently, posing potential risks and losses.

The sample code is illustrated in. Fig. 4. In this example, the function kill is a public function that directly invokes SELFDESTRUCT to terminate the contract and transfers the balance to the caller's address (i.e., 'msg.sender').

To mitigate this vulnerability, stringent permission controls can be implemented for sensitive functions, allowing only authorized addresses to invoke operations that may lead to self-destruction. Furthermore, Solidity's modifiers can be utilized to ensure that SELFDESTRUCT operation occurs only under specific conditions.

```solidity
1 contract Suicidal {
2   function kill() public {
3       selfdestruct(msg.sender);
4       //Vulnerability Location
5   }
6 }
```

**Fig. 4.** Example Code of The Unprotected SELFDESTRUCT Vulnerability

## 4. Methodology

### 4.1 Problem Description

Current approaches to smart contract security primarily focus on static analysis, examining contracts at specific points in time or targeting particular vulnerabilities. This narrow focus often overlooks the dynamic evolution of vulnerabilities throughout the contract's lifecycle, especially during transaction processes. There is a clear need for deeper exploration of how vulnerabilities manifest and evolve during different lifecycle stages, particularly those involving complex operations like DELEGATECALL and SELFDESTRUCT. This study aims to address this gap by analyzing the dynamic characteristics and developmental trends of vulnerable smart contracts, providing insights for more effective security measures and risk mitigation in the evolving Ethereum ecosystem.

### 4.2 Data Collection

Initially, our experiment needs efficient and accurate vulnerability detection tools along with a set of vulnerable smart contracts. Starting with the SmartBugs wild dataset [43], which contains code and addresses for over 47,000 smart contracts. Based on the vulnerability detection and lifecycle stage identification proposed in Section 3, we conduct a preliminary vulnerability assessment on 500 randomly selected smart contracts using an integrated tool incorporating exploit code from SmartBugs2 for detection. Results show that four tools, namely Mythril [44], Confuzzius [45], Slither [46], and Sfuzz [47], exhibit higher detection rates across multiple vulnerability types with increased accuracy, particularly during deployment and destruction stages. Hence, these tools are selected for smart contract vulnerability detection. To enhance confidence in our findings, if two or more tools identify the same vulnerability in a given smart contract, we assert the presence of a genuine vulnerability in that contract [43]. Finally, we employ these tools to identify 800 smart contracts with distinct-stage vulnerabilities in a random subset of 5,000 contracts.

Subsequently, gathering a diverse array of relevant transactional information is crucial. To retrieve all transactions associated with identified vulnerable contracts (both internal and external), a web crawler is developed to obtain their complete transaction history from the Etherscan website [33]. Additionally, to construct the ego network, we collect the entire transaction history of contracts interacting with the identified vulnerable contracts. This dataset encompasses over ***1.54 million*** transaction records, offering a holistic view of interactions among these contracts. The fields and their meanings are detailed in Table 1. From this dataset, comprehensive transaction-related information, including transaction counts, amounts, and transaction neighbors within specific time intervals, is available.

Table 1. Primary Fields of Ethereum Transaction Data

| Field | Description |
|---|---|
| Contract Address | The contract address |
| Method | The function you can call to perform specific tasks |
| Date | The date and time when the transaction occurred |
| Transaction Hash | The unique identifier for a transaction on the blockchain |
| Block | The block number |
| From | The address of the transaction initiator (20-byte string) |
| To | The address of the transaction recipient (20-byte string) |
| Value (ETH) | The transaction amount in Ether (ETH) |

### 4.3 Data Processing and Network Construction

In our study, vulnerable smart contracts are treated as nodes, with addresses directly involved in transactions designated as first-order transaction neighbors. To improve the relevance and accuracy, we remove irrelevant records, such as self-loop transactions, where a smart contract interacts with itself, and failed transactions from the dataset.

We model each vulnerable smart contract as a node, denoted by $G = (V, E)$, where $V$ includes the smart contract EGO and its neighboring ALTERs involved in transactions. Transaction records consist of marked smart contract transactions and their first-order transaction neighbors. Edges, represented as $e = (f, t)$, f designates from, representing the address of the transaction initiator, while t designates to, representing the address of the transaction recipient. The symbol e signifies asset transfers from one address to another, thereby establishing a directed multigraph for each transaction ego network.

## 5. Features of Smart Contracts

In this section, we focus on and systematically examine various types of features associated with smart contracts, collecting statistical data for analysis.

### 5.1 Transaction Features

Transaction attributes can most directly reflect the behavioral characteristics of Ethereum accounts. For instance, the lifecycle length of a smart contract refers to the lifespan from its creation to termination, offering insights into the contract's stability and long-term availability. The number of neighbors for a smart contract indicates the number of other Contract Accounts or Externally Owned Accounts (EOAs) interacting with it, reflecting the degree of association and interoperability.

### 5.1.1 Lifespan of Contracts

We retrieve data from the Etherscan website, specifically documenting the monthly count of smart contracts created [33]. In the Ethereum system, thousands of new accounts are created daily, exhibiting significant variation in transaction frequency. The rapid growth in smart contracts is accompanied by an increase in security issues, highlighting the need for more effective methods to analyze critical security concerns and the developmental patterns of smart contracts.

We define the lifespan of a contract as the time interval between its first and last transactions (until the data was collected). For each contract type, we compute the average, median, and standard deviation of its lifespan, as shown in **Table 2**, and depict the distribution in **Fig. 5**. Contracts with vulnerabilities during both the deployment and execution stages are categorized as having other vulnerabilities.

Table 2. Statistics for The Lifespan of Each Stage (Days)

| Label | Others | Delegatecall | Selfdestruct |
|---|---|---|---|
| Mean | 340.48 | 226.84 | 77.83 |
| Median | 33.85 | 3.60 | 0.84 |
| Std. | 569.31 | 401.87 | 234.10 |

Smart contracts with different vulnerabilities exhibit diverse lifespan distributions. As shown in **Table 2** and **Fig. 5**, over 90% of smart contracts with selfdestruct vulnerabilities have a lifespan of 100 days or less. About 60% of contracts with other vulnerabilities also last within 100 days, though approximately 10% persist beyond 1400 days. For contracts with delegatecall vulnerabilities, around 70% last within 100 days, and all remain within the 1400-day range, indicating a more uniform distribution compared to other vulnerabilities.

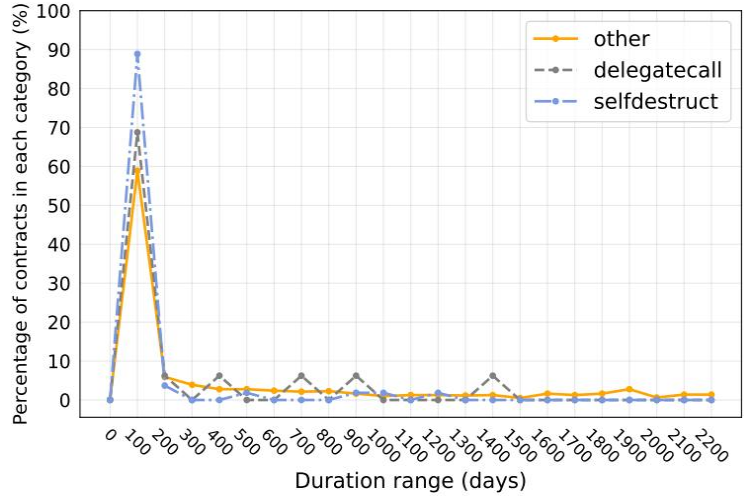

**Fig. 5.** Contract Lifespan Duration Distribution

The variation in lifespan distribution can be attributed to the distinct nature of vulnerabilities. Specifically, the median lifespan of smart contracts with other vulnerabilities is only 33.85 days, but they exhibit the highest standard deviation, indicating some contracts with exceptionally long lifespans. Notably, only contracts with vulnerabilities during the deployment and execution stages exceed 1400 days. To further investigate these long-lifespan contracts, we examine the top eight vulnerabilities with the proportion of contracts with extended lifespans, as shown in **Fig. 5**. As seen in **Fig. 6**, 88% of vulnerabilities pose minimal risk, requiring only functional enhancements to the smart contracts. Only Assert violation, Shadowing abstract, and Integer overflow/underflow vulnerabilities present high risks [46]. This analysis reveals that even high-risk vulnerabilities are seldom exploited over extended periods [48], while low-risk vulnerabilities are generally disregarded, allowing these contracts to persist on the blockchain and remain operational for prolonged durations.

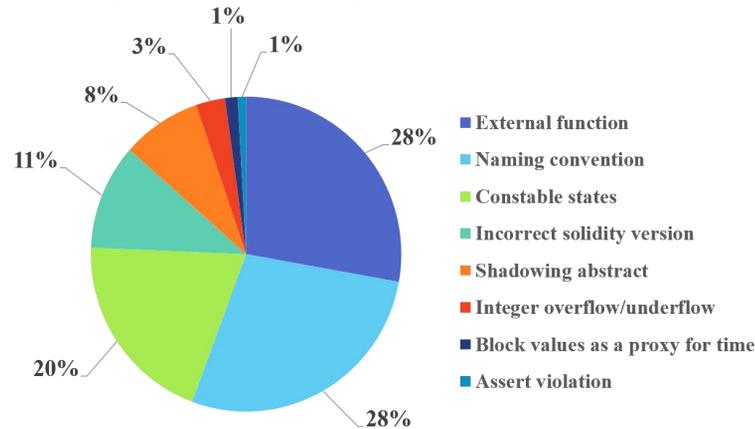

**Fig. 6.** Top Eight Vulnerabilities in Smart Contracts Which Have Lifespans Exceeding 1400 Days

To comprehend the transactional characteristics of smart contracts with vulnerabilities at different periods of lifespan, we employ a variable-length time window. Distinguished from prior studies [35, 49], our approach divides the lifespan, from the first transaction to the last, of each contract into five distinct periods (P1~P5) for analysis. This partitioning yields a clear chronological perspective, aligning more closely with the objectives of our study.

**Finding 1:** Smart contracts with selfdestruct vulnerabilities exhibit a lifespan primarily within 100 days, displaying a remarkably uniform distribution. In contrast, smart contracts with other vulnerabilities have lifespans over 1400 days.

**Finding 2:** Despite the heightened risks inherent in some smart contracts, it is not necessarily the case that these vulnerabilities will be maliciously exploited. As a result, these smart contracts have a significant lifespan.

### 5.1.2 Changes in The Number of Transactions

We assess transaction volume as an indicator of smart contract activity throughout its lifespan. To account for the uneven distribution of contract types, we present the proportional representation of transaction volumes across periods in **Fig. 7**, which indicates that, collectively, smart contracts with vulnerabilities (excluding selfdestruct vulnerabilities) contribute around 80% of the total transaction volume initially, followed by a gradual decline. In contrast, contracts with delegatecall vulnerabilities exhibit fluctuating transaction volumes, characterized by phases of decrease, increase, and subsequent decrease.

These smart contracts initially exhibit a higher number of transactions due to extensive user interactions following deployment. We conduct an in-depth analysis of smart contracts with delegatecall vulnerabilities, which show fluctuating transaction volumes. Our findings indicate that these contracts are primarily involved in processes like voting through proxy calls and decentralized lending. In contrast, transactions for smart contracts with selfdestruct vulnerabilities and those with other vulnerabilities in p3, p4, and p5 are nearly negligible.

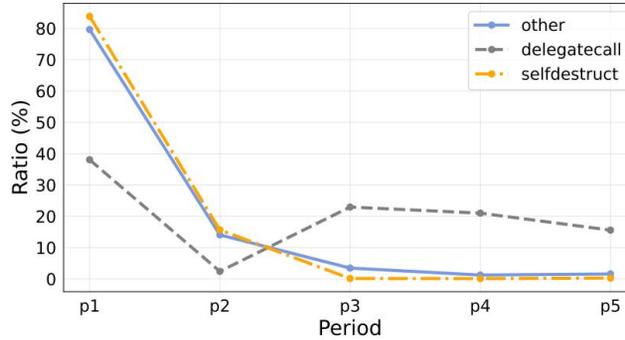

**Fig .7.** Ratio of Transactions for Each Type of Smart Contract per Period

It is noteworthy that transactions between smart contracts are directional, involving both incoming and outgoing transfers. In the former case, contracts receive Ether, while in the latter, Ether is sent to others. To analyze these changes, we introduce an indicator $\alpha$, calculated as the ratio of incoming to outgoing transactions, considering only non-zero amounts. The variations in $\alpha$ are shown in **Table 3**. When $\alpha$ exceeds 1, it indicates a higher frequency of incoming transactions, while $\alpha$ less than 1 suggests more outgoing transactions. An $\alpha$ value of 0 signifies no incoming or outgoing transactions.

**Table 3.** Changes in Metric $\alpha$

| Period | P1 | P2 | P3 | P4 | P5 |
|---|---|---|---|---|---|
| **Other** | 0.72 | 0.31 ↓ | 2.38 ↑ | 0.63 ↓ | 1.19 ↑ |
| **Delegatecall** | 0.00 | 2.00 ↑ | 2.00 | 2.00 | 0.67 ↓ |
| **Selfdestruct** | 1.77 | 1.92 ↑ | 0.35 ↓ | 0.63 ↑ | 0.34 ↓ |

The performance of $\alpha$ varies across different categories of smart contracts. Contracts with delegatecall vulnerabilities involve transactions with zero values, a result of DELEGATECALL's nature, where the initiating contract delegates execution without transferring funds. This is common in proxy contracts, where users invoke functions in the target contract, preserving the caller's context, including 'msg.sender' and 'msg.value'. This allows attackers to exploit the caller's context or affect the calling contract's storage [50].

Smart contracts with selfdestruct vulnerabilities show a significantly higher number of outgoing transactions in their final lifespan period compared to incoming transactions. A manual inspection of outgoing transactions in P5 revealed that 58.5% of these contracts self-destructed after transferring assets, as confirmed by reviewing their transaction history on Etherscan.

**Finding 3:** Smart contracts with delegatecall vulnerabilities exhibit notable fluctuation in transaction numbers throughout their entire lifespan, while those with other vulnerabilities and selfdestruct vulnerabilities both display a gradual decrease in the transaction numbers.

**Finding 4:** In P5 phase, the majority of contracts containing selfdestruct vulnerabilities selfdestructed after transferring assets.

### 5.1.3 Changes in the Amount of Transactions

Unlike social networks, edges between nodes in the Ethereum transaction network carry monetary information, offering deeper insights into smart contract activities. **Fig. 8** illustrates the transaction amounts for various smart contracts across different time intervals.

**Fig. 8** demonstrates that smart contracts with other vulnerabilities, as well as those with selfdestruct vulnerabilities, experience incoming transaction amounts exceeding outgoing transactions during the initial phase of their lifespan. In contrast, smart contracts with delegatecall vulnerabilities exhibit consistently low transaction amounts. Smart contracts with other vulnerabilities and selfdestruct vulnerabilities demonstrate substantially lower transaction amounts in the P3, P4, and P5 phases compared to the P1 phase.

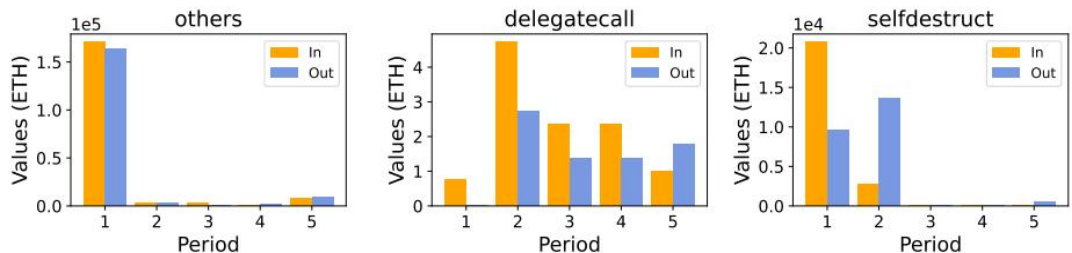

**Fig. 8.** Changes of The Transaction Amount per Period

**Finding 5:** Smart contracts with selfdestruct and other vulnerabilities demonstrate a notable decline in transaction amounts during the P3, P4, and P5 phases.

**Finding 6:** Smart contracts with delegatecall vulnerabilities typically manifest in transactions with notably low amounts, in contrast to those with selfdestruct and other vulnerabilities, which tend to involve higher transaction amounts.

**Finding 7:** Smart contracts with other vulnerabilities have transaction amounts in the first phase that account for approximately 90% of the entire lifespan.

### 5.1.4 Changes in the Number of Neighbors

In the Ethereum ecosystem, individuals typically remain anonymous to each other. Nevertheless, insights into the social nature of Ethereum contracts can be gleaned by comparing the number of transaction partners between different Ethereum contracts. A contract can engage in multiple transactions with another contract, termed as transaction neighbors. **Fig. 9** and **Fig. 10** illustrate the variations in the number of incoming and outgoing transaction neighbors. Moreover, the trans_in represents incoming neighbors, and the trans_out represents outgoing neighbors.

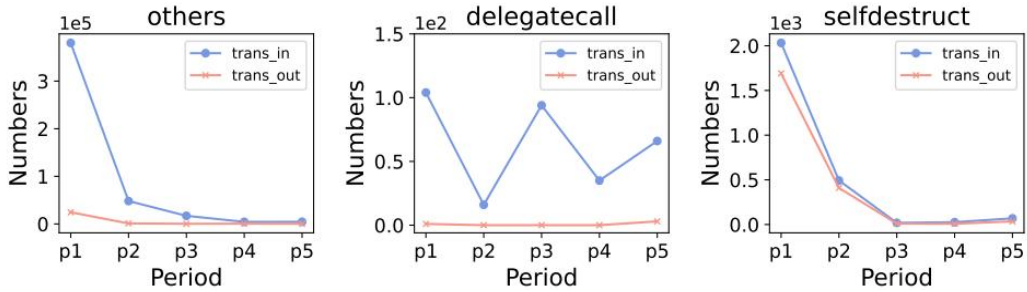

**Fig. 9.** Change in Incoming and Outgoing Neighbors per Period (Transaction Value Can Be 0)

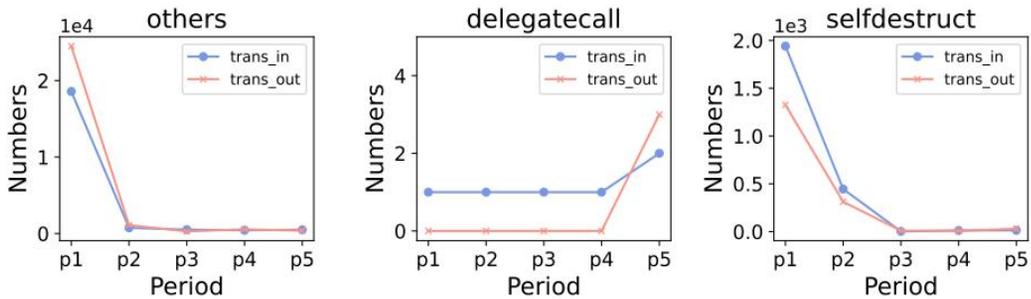

**Fig. 10.** Change in Incoming and Outgoing Neighbors per Period (Transaction Value Can Not Be 0)

From these two figures, we gain unique insights into the transactional neighbors. Contracts with delegatecall vulnerabilities and those with other vulnerabilities often transact with zero amounts among their neighbors. Delegatecall contracts primarily involve zero-value transactions, consistent with their role as invoked contracts focused on data storage or function calls, supporting the findings in sections 5.1.2 and 5.1.3. In contrast, smart contracts with selfdestruct vulnerabilities show nearly equal incoming and outgoing neighbors, with most transactions involving non-zero amounts. Excluding delegatecall contracts, both types of contracts exhibit a gradual reduction in incoming and outgoing neighbors.

**Finding 8:** If the transaction amounts can be zero, contracts with delegatecall vulnerabilities demonstrate a consistent number of outgoing neighbors.

**Finding 9:** Regardless of whether transaction amounts can be zero, contracts with selfdestruct vulnerabilities exhibit an equivalent quantity of incoming and outgoing neighbors.

### 5.1.5 Changes in the Number of Old and New Neighbors

Fixed interaction partners significantly characterize an Ethereum contract. In each contract lifespan, the first appearance of neighboring contracts is termed new neighbors, while previously encountered ones are labeled old neighbors. **Fig. 11** and **Fig. 12** illustrate the proportions of new and old neighbors among transaction partners, providing insights into contract transaction tendencies. Each bar in the figure is divided into non-overlapping blue and yellow segments, representing quantities of new and old neighbors in each stage.

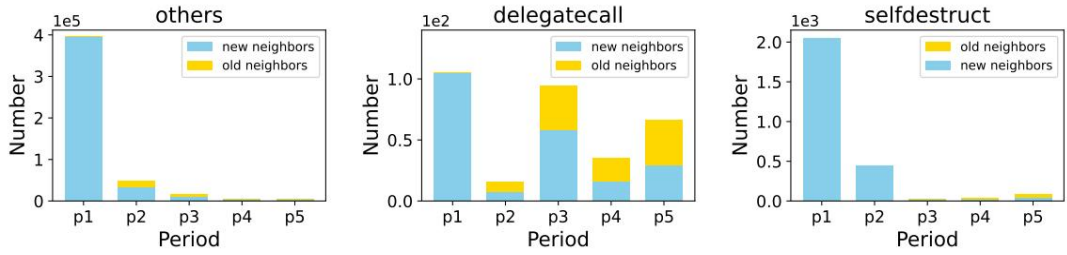

**Fig. 11** Changes in New and Old Neighbors per Period(Transaction Value Can Be Zero)

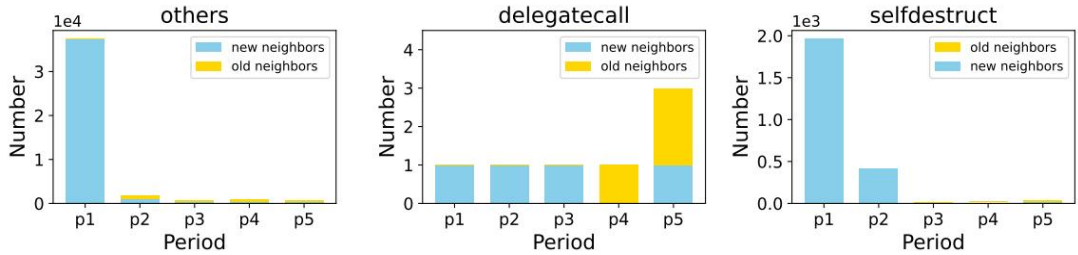

**Fig. 12** Changes in New and Old Neighbors per Period(Transaction Value Can Not Be Zero)

Analyzing the graph, we observe that over 90% of smart contracts with selfdestruct vulnerabilities interact with new neighbors. This aligns with our earlier analysis, where new neighbors quickly initiate outgoing transactions after incoming ones in the current phase. Notably, in the third phase of contracts with selfdestruct vulnerabilities, there is no fund transfer, but in the final phase, new neighbors engage in minimal fund transfers. This reaffirms our earlier finding that, despite the prevalence of vulnerabilities, the malicious exploitation rate remains low, consistent with Perez et al. [48]. In contrast, smart contracts with delegatecall vulnerabilities exhibit the highest ratio of old neighbors, while contracts with other vulnerabilities consistently engage with a portion of old neighbors, regardless of whether transaction amounts are zero.

**Finding 10:** Regardless of whether transaction amounts can be zero, smart contracts with delegatecall vulnerabilities have the highest ratio of old neighbors.

**Finding 11:** Over 90% of smart contracts with selfdestruct and other vulnerabilities interact with new neighbors, and these neighbors promptly initiate outgoing transactions after incoming transactions during the current phase.

### 5.2 Properties of Transaction Ego Network

In addition to transaction features, the structural attributes of Ethereum contracts are crucial for observing transaction patterns. By exploring network properties such as network density and local clustering coefficient, we can better understand the neighborhood complexity of flagged contracts. Given the minimal impact of multiple edges on detecting network characteristics, we analyze the simple directed graph version of ego networks in this section.

### 5.2.1 Temporal Variation of Network Density

The variation in network density not only signifies the sustained trends within the network but also implies the degree of connectivity among all nodes in the network. For an ego network, its density can be defined as:

$$d = \frac{|E|}{(|V| - 1) * |V|} \qquad (1)$$

Here, $E$ represents the number of directed edges, $V$ denotes the number of nodes in the ego network graph, and $|\cdot|$ denotes the operation for computing set size.

Consequently, we calculate the average network density for each type of smart contract at different periods. As shown in **Fig. 13**, the average ego network density for all smart contracts is consistently below 0.5. However, apart from smart contracts with other vulnerabilities, the manifestations of the remaining two categories do not align entirely with those reported by Zhao et al. [51]. It is evident that, with the evolution of Ethereum, an increasing number of nodes appear in the transaction network of smart contracts. Yet, since most nodes do not engage in transactions with each other, it leads to network sparsity. Therefore, the ego network density of smart contracts with other vulnerabilities aligns with the regularity observed in network development.

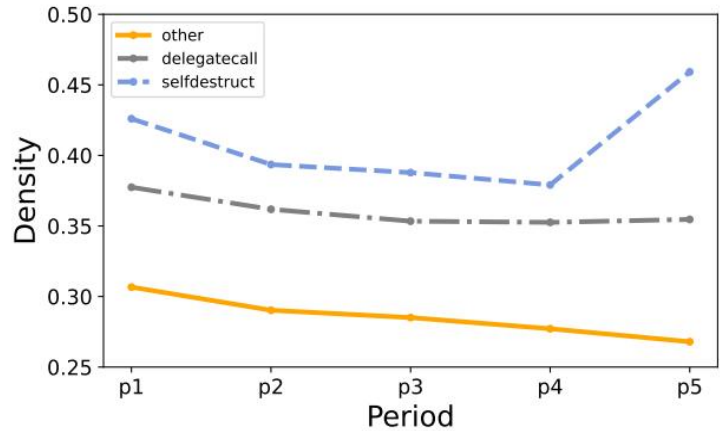

**Fig. 13.** The Density of Ego Network per Period

We aim to comprehend the reasons behind the discrepancy in ego network density and blockchain network development patterns observed in the other two categories of smart contracts. The smart contract with the highest network density is the one containing the selfdestruct vulnerability, and its density notably differs from that of smart contracts with other vulnerabilities. As detailed in Sections 5.1.4 and 5.1.5, smart contracts with the selfdestruct vulnerability exhibit fewer neighbors, with almost no addition of new neighbors in the last three phases. This contributes to the increase in ego network density in the final phase. Similarly, smart contracts with the delegatecall vulnerability also demonstrate a higher ego network density than those with other vulnerabilities. As evidenced in Section 5.1.5, this type of contract has the highest ratio of old neighbors and the lowest number of neighbors overall. Therefore, we deduce that the number of neighbors (i.e., node counts) is a crucial factor influencing network density, aligning with the findings of Wu et al. [20].

**Finding 12:** We consider that the node count within the subnet is a key factor influencing network density.

### 5.2.2 Local Clustering Coefficient

The local clustering coefficient is a commonly used metric to evaluate tight relationships among neighbors within a cluster, offering insights into the inherent sparsity of Ethereum's local structure. It is calculated by dividing the number of connections between nodes in a node's neighborhood by the potential number of connections among them. Thus, in an ego network, the local clustering coefficient is defined as:

$$c = \frac{|\{e_{uv}: u, v \in N_i, e_{uv} \in E\}|}{|N_i| * (|N_i| - 1)} \quad (2)$$

Here, $u$ and $v$ represent distinct non-central neighbor nodes in the ego network. $e_{uv}$ signifies the directed transaction edge from node $u$ to node $v$ within the ego network. $N_i$ represents the first-order neighbors of the ego network, and $|\cdot|$ denotes the operation for computing set size. Furthermore, we calculate the average local clustering coefficient $C_{avr}$.

The results in **Table 4** highlight that smart contracts with selfdestruct vulnerabilities exhibit the highest clustering coefficient of 0.015 among all account types, implying a 1.5% probability of transactions between these contracts, indicating their closer interconnectedness. **Fig. 14** supports this, suggesting higher network density due to the relatively fewer neighbors of these contracts. Upon closer inspection, these contracts primarily facilitate functions such as asset storage, retrieval, auctions, and gaming.

**Table 4.** The Average Clustering Coefficient of Various Types of Smart Contracts

| Label | Others | Delegatecall | Selfdestruct |
|---|---|---|---|
| $C_{avr}$ | 0.007 | 0.001 | 0.015 |

Smart contracts with delegatecall vulnerabilities exhibit the smallest coefficient, suggesting weaker relationships among them. However, in contrast to the preceding figure, a non-negligible network density is observed. This implies that network density is not solely influenced by the clustering degree. The increased density in these networks is attributed to fewer neighbors and a higher ratio of old neighbors. Thus, network size, clustering degree, and the proportion of old neighbors collectively determine overall network density.

**Finding 13:** Smart contracts with selfdestruct vulnerabilities exhibit the highest clustering coefficient, and both the clustering coefficient and the proportion of old neighbors are additional factors determining network density.

## 6. Detection Experiment

Through the preceding analysis in Section 5, we have identified distinct dynamic transaction behaviors and transaction neighborhood characteristics among different smart contracts. Consequently, by utilizing machine learning classification models on the contract features we designed, we can ascertain the alignment of our findings with general characteristics and determine the extent to which these smart contracts can be distinctly differentiated.

### 6.1 Experiment Settings

We utilize the seven foundational features analyzed in Section 5. Five widely employed machine learning models: Logistic Regression (LR), Random Forest (RF), Support Vector Machine (SVM), Decision Tree (DR), and K-Nearest Neighbors (KNN) are selected for blockchain contracts classification. Default parameters from the *sklearn* library are used for each model. The sample set is split into 60% training, 30% testing, and 10% validation subsets. Evaluation includes metrics like accuracy, precision, recall, and F1 score, with validation utilizing cross-validation.

## 6.2 Experiment Results

Table 5 shows that KNN classification is notably effective, achieving a precision of 0.715 and a recall of 0.726. Fig. 14 presents the Receiver Operating Characteristic (ROC) curve generated from the KNN model's predictions, illustrating the trade-off between the true positive rate (sensitivity) and the false positive rate (1-specificity) across different thresholds. The area under the curve (AUC) is 0.7222, reflecting the model's ability to discriminate between classes. We also present confusion matrices for the five classification models.

**Table 5.** Experiment Results With Different Classification Models

| Model | Accuracy | Precision | Recall | F1 score |
|---|---|---|---|---|
| LR | 0.589 | 0.612 | 0.589 | 0.594 |
| SVM | 0.560 | 0.560 | 0.560 | 0.561 |
| DR | 0.305 | 0.331 | 0.305 | 0.268 |
| RF | 0.344 | 0.597 | 0.344 | 0.196 |
| KNN | **0.726** | **0.715** | **0.726** | **0.713** |

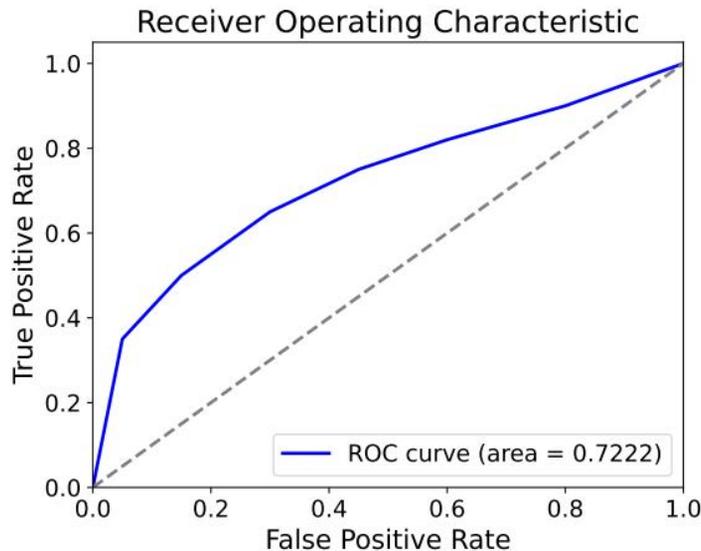

**Fig. 14.** ROC Curve for KNN Model Predicting Result

A confusion matrix is a tabular representation for evaluating classification performance, with percentages indicating the proportions of each class in the dataset. Shading in the matrix provides a quick visual of the model's accuracy, where darker shades denote higher values. The confusion matrix in Fig. 15, based on the KNN model, highlights distinctions among smart contracts but also reveals misclassifications of contracts with other vulnerabilities as different types. This may stem from the diverse vulnerabilities within these contracts, where some share features with selfdestruct and delegatecall vulnerabilities.

**Finding 14:** Significant distinctions indeed exist among smart contracts that contain vulnerabilities at different lifespan stages. And the features we designed can effectively differentiate these smart contracts.

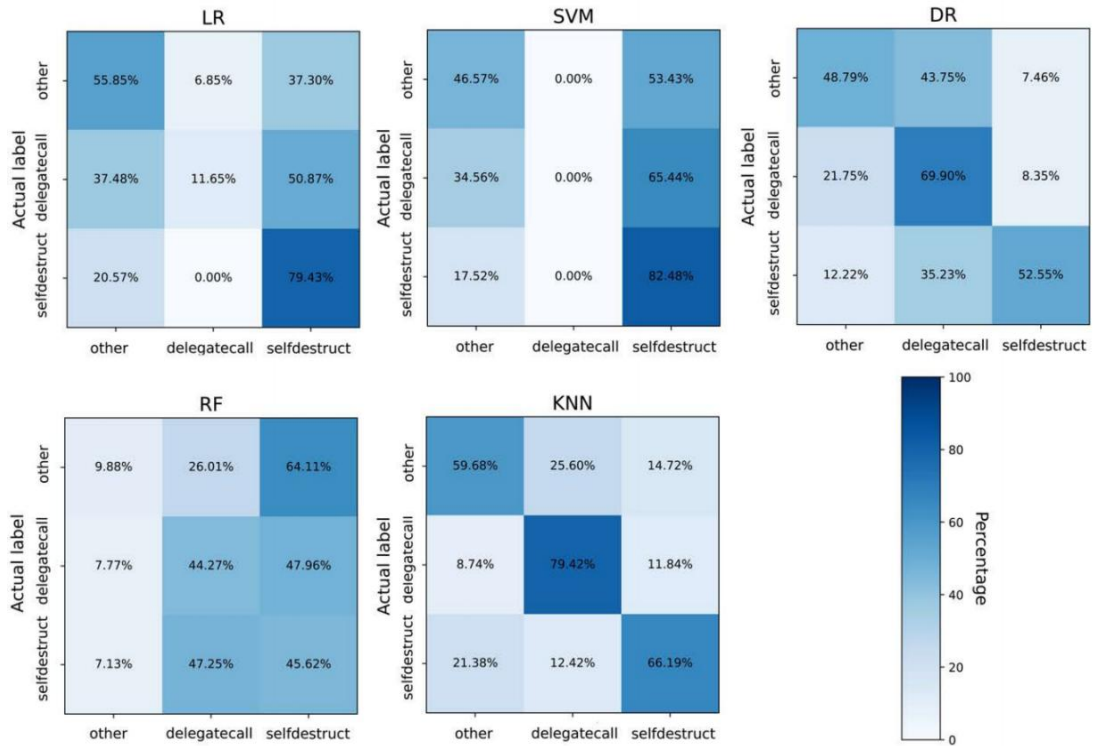
**Fig. 15.** Confusion Matrix of The Five Model Predictions

## Discussion

Our study addresses a gap in smart contract vulnerability research by conducting a comprehensive lifecycle analysis of vulnerable contracts, unlike prior work focused mainly on static analyses or specific vulnerability types. By examining transaction dynamics and ego network features, we provide deeper insights into the characteristics of vulnerable contracts. Notably, most vulnerable contracts have lifecycles under 100 days, with selfdestruct vulnerabilities being particularly short-lived. In contrast, over 95% of transactions involving contracts with delegatecall vulnerabilities lack asset transfers, reflecting their role in state or function calls.

Our findings reveal distinct characteristics among various vulnerability types, leading to the development of two feature categories: fundamental and dynamic differentiating features. The KNN model achieved superior classification performance, though some confusion between vulnerabilities arose, likely due to overlapping feature manifestations. Despite limitations, including dataset specificity and sample size, our results provide valuable insights into Ethereum smart contract development and trends.

## 8. Related Work

### 8.1 Ethereum Analysis

The current analysis of Ethereum can be categorized into two approaches: comprehensive analysis of the entire Ethereum network and focused analysis of specific aspects [52].

Comprehensive analyses of the Ethereum network have increasingly focused on transaction networks. Lee et al. [53] conducted an extensive study using network analysis methods to explore interactions between users and smart contracts across four Ethereum networks and three token sub-networks. Their findings revealed that while Ethereum networks differ from social networks due to low transitivity and the absence of community structures, they still exhibit small-world properties and strong connectivity. Building on this, Lin et al. [54] introduced the temporal weighted multidigraph model, incorporating temporal and monetary attributes to improve transaction analysis accuracy.

Guo et al. [55], Li et al. [56], and Lin et al. [57] have focused on analyzing transaction relationships. Guo et al. [55] employed a framework based on network science, identifying a heavy-tailed distribution in Ethereum transaction characteristics, which aligns with power-law functions. This analysis also showed that popular nodes often connect with numerous ordinary users, challenging the existence of a rich club phenomenon. Building on this, Li et al. [56] proposed TTAGN, a method for detecting phishing fraud using temporal relationships in transaction data. Their approach, integrating graph neural networks and transaction statistics, improves fraud detection on Ethereum. Lin et al. [57] explored Ethereum's transaction relationship evolution, constructing a micro-level network model that uncovered a star-shaped structure and highlighted the role of transaction frequency in network evolution, introducing a link prediction method to forecast future relationships.

Furthermore, the current analysis of Ethereum extends to various aspects. Yaish et al. [58] analyzed Ethereum's consensus mechanism, focusing on an attack method called Uncle Maker within the proof-of-work (PoW) consensus. Their findings revealed that Ethereum's second-largest mining pool executed this attack for over two years. Bai et al. [59] examined transaction pattern evolution on Ethereum using temporal graph analysis. They constructed User-to-User (UUG), Contract-to-Contract (CCG), and User-to-Contract (UCG) graphs, validating transaction fairness and wealth distribution while highlighting the growing importance of smart contracts. Their study also explored spatiotemporal dynamics, uncovering a correlation between transaction patterns and Ether price fluctuations.

### 8.2 Smart Contract Analysis

Current smart contract analyses mainly target vulnerability detection [40, 60–62], with various approaches developed to improve accuracy and efficiency. Ethainter, introduced by Brent et al. [40] detects composite attacks by analyzing information flow and data sanitization within smart contracts, achieving a precision of 82.5%. Zhang et al. [60] proposed MODNN, a scalable neural network model capable of detecting multiple vulnerabilities simultaneously, with an average F1 score of 94.8%. Sendner et al. [61] developed ESCORT, a deep learning-based tool using transfer learning to detect new vulnerability types with an average F1 score of 96%. Wu et al. [62] introduced Peculiar, a model leveraging pre-training on crucial data flow graphs, achieving 91.8% precision and 92.4% recall in reentrancy vulnerability detection, outperforming existing methods.

What's more, current analyses of smart contracts increasingly focus on gas consumption. Chen et al. [63] introduced GasChecker, a tool using symbolic execution and MapReduce to identify gas-inefficient programming patterns. They summarized ten such patterns and proposed the FBLB strategy to improve resource efficiency. Similarly, Albert et al. [64] developed Gasol, which provides cost models for analyzing and optimizing gas usage in Ethereum smart contracts, assisting in detecting resource-oriented attacks.

In contrast, Hu et al. [65] focused on detecting fraudulent activities in smart contracts through SCSGuard, a deep learning-based framework. This tool leverages bytecode features and an attention mechanism based on GRU networks to identify fraud. From a broader perspective, Oliva et al. [66] analyzed smart contract activity, categorization, and code complexity on Ethereum, revealing a concentration of activity in a small subset of contracts. Despite the interest in blockchain applications, the primary use of smart contracts remains in token management, particularly for ICOs and crowdfunding.

## 9. Conclusion and Future Perspectives

Our study offers a comprehensive analysis of smart contract vulnerabilities by categorizing them into three distinct lifecycle stages: deployment and execution, upgrade, and destruction. By identifying features associated with vulnerabilities at each stage, we enhance detection capabilities and deepen the understanding of security personnel. These features are also effectively applied in machine learning-based approaches for vulnerability detection, demonstrating their utility throughout a smart contract's lifecycle. Furthermore, we highlight the challenge of distinguishing between contracts with different vulnerabilities, particularly those with selfdestruct or delegatecall issues.

However, there are misclassifications among various contracts. To address these misclassifications, future research should focus on examining whether the confusion arises from internal or external transaction factors. Additionally, expanding the dataset to include a larger and more current collection of smart contracts will be crucial for refining classification models and ensuring the generalizability of our findings.